\documentclass[10pt]{article}
\usepackage[dvips]{graphicx}
\def\obrxxi{\unitlength=1bp\begin{picture}(220,180)(0,-15)
\put(10,0){\makebox(0,0)[bl]{\includegraphics[width=0.43
\textwidth]{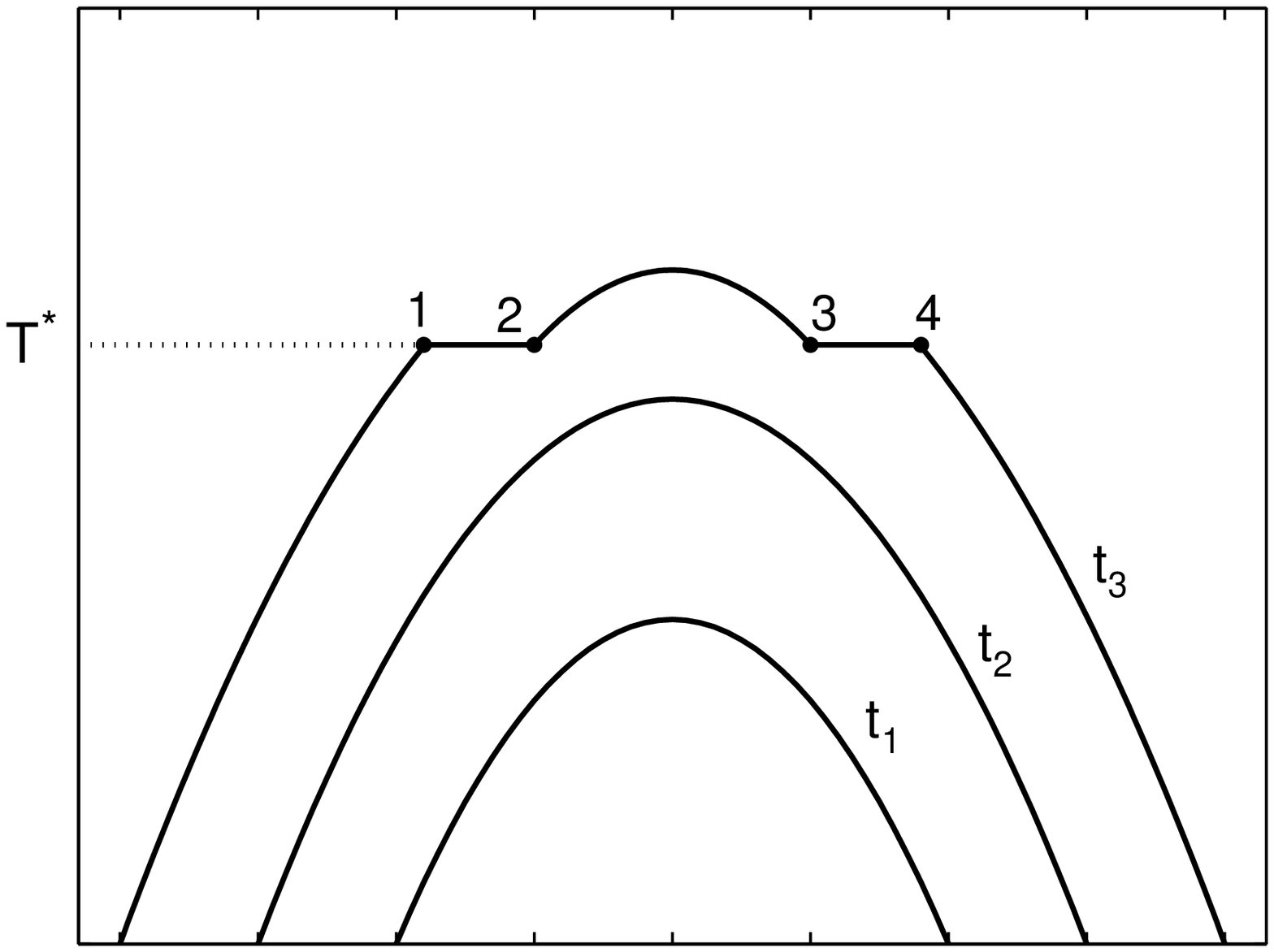}}} \put(80,120){\makebox(0,0)[ct]{$T(x)$}}
\put(105,-5){\makebox(0,0)[ct]{$x$}}
\put(20,-15){\makebox(0,0)[ct]{$a)$}}
\put(205,-35){\makebox(0,0)[ct]{
\begin{tabular}[t]{p{123mm}}
{\bf Fig.~1.} Evolution of the temperature distribution  in the
pattern for: a) a case of spatially distributed  sources of heat,
b) the classical Stefan problem (with heating from left to right).
Here $t_1 < t_2 < t_3 .$
\end{tabular}}}
\end{picture}}

\def\obrxxii{\unitlength=1bp\begin{picture}(220,180)(0,-15)
\put(10,0){\makebox(0,0)[bl]{\includegraphics[width=0.43
\textwidth]{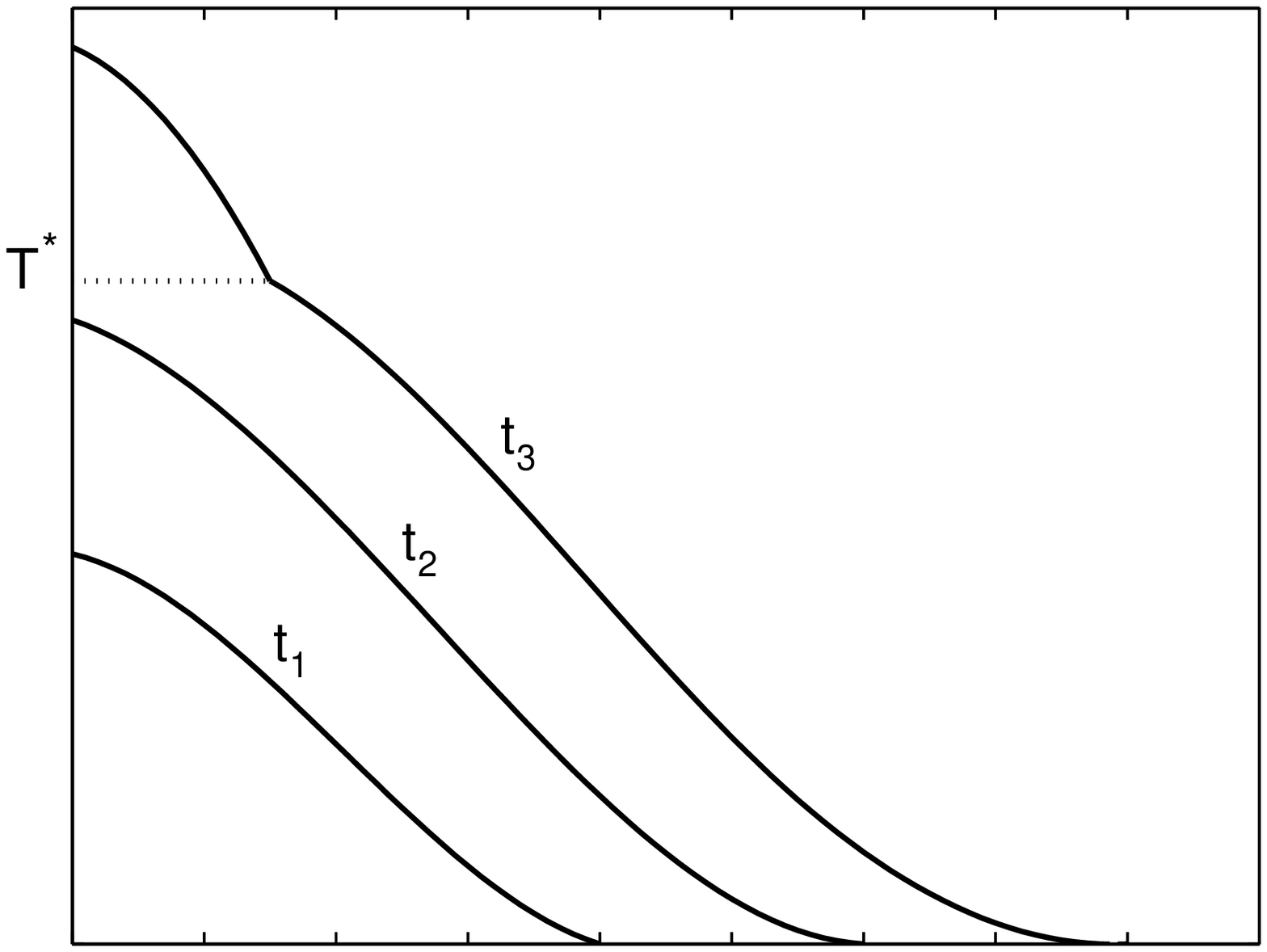}}} \put(100,120){\makebox(0,0)[ct]{$T(x)$}}
\put(105,-5){\makebox(0,0)[ct]{$x$}}
\put(20,-15){\makebox(0,0)[ct]{$b)$}}
\end{picture}}

\def\obri{\unitlength=1bp\begin{picture}(220,190)(0,-15)
\put(10,0){\makebox(0,0)[bl]{\includegraphics[width=0.50
\textwidth]{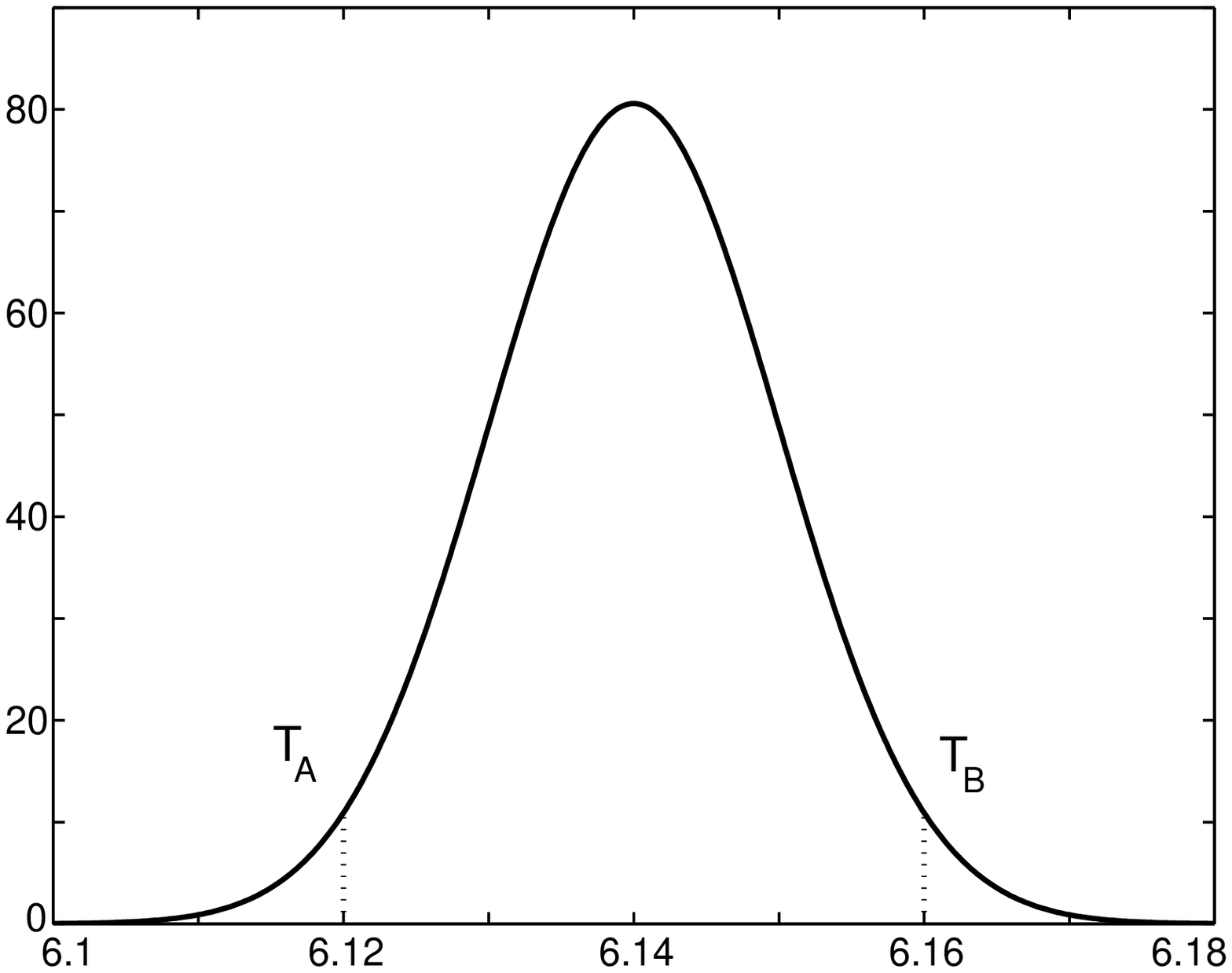}}} \put(55,150){\makebox(0,0)[ct]{
$\delta(T-T^*, \Delta)$}} \put(105,-5){\makebox(0,0)[ct]{$T$}}
\put(115,-20){\makebox(0,0)[ct]{
\begin{tabular}[t]{p{75mm}}
{\bf Fig.~4.} Approximate $\delta$-function with maximum at $T^*$
and smearing equals $\Delta$ \hspace {1mm}(for $\Delta=0,03$)
\end{tabular}}}
\end{picture}}

\def\obrii{\unitlength=1bp\begin{picture}(220,190)(0,-15)
\put(10,0){\makebox(0,0)[bl]{\includegraphics[width=0.43
\textwidth]{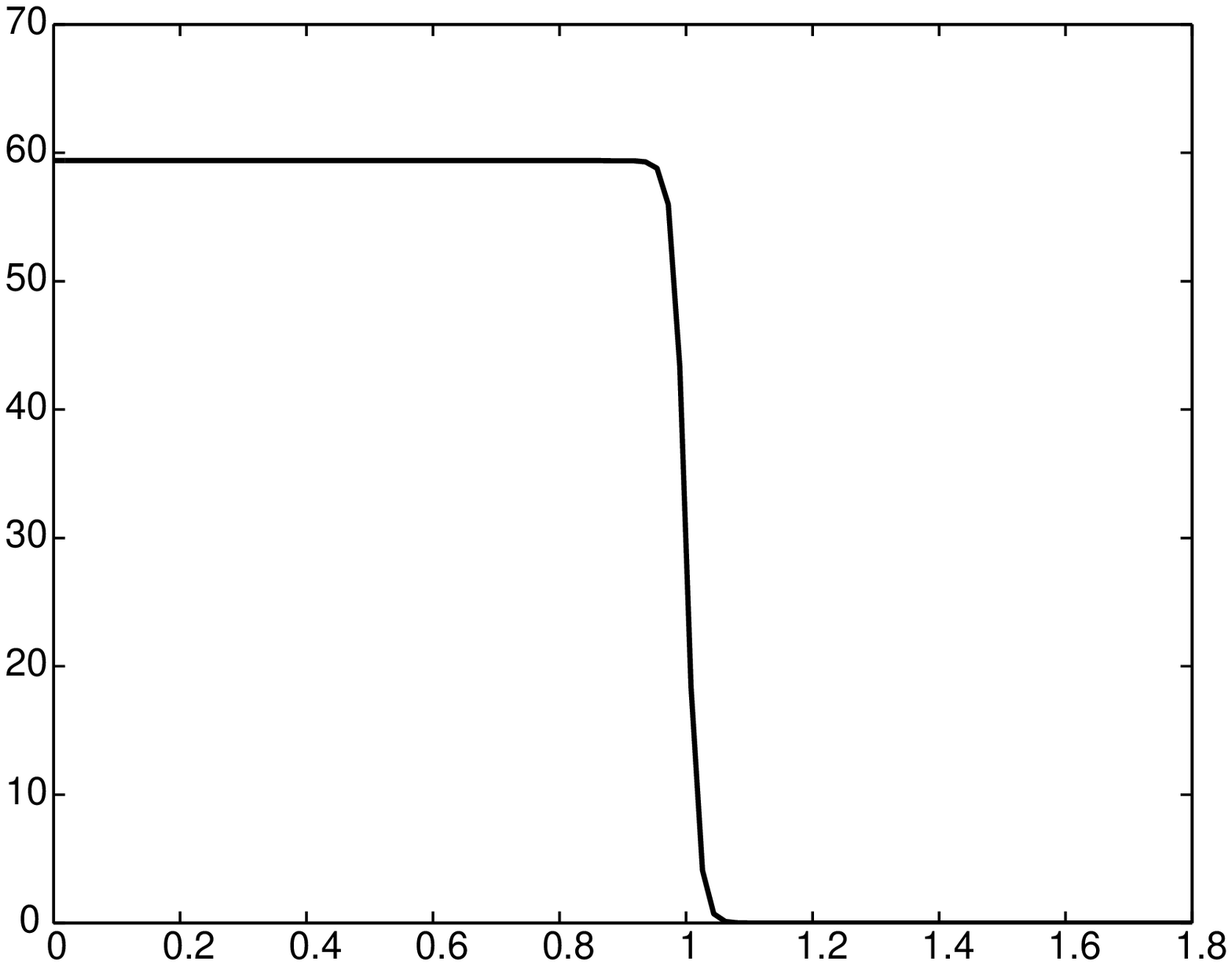}}}
\put(50,150){\makebox(0,0)[ct]{$q(x=0,t)$}}
\put(105,-5){\makebox(0,0)[ct]{$t$}}
\put(95,-20){\makebox(0,0)[ct]{
\begin{tabular}[t]{p{63mm}}
{\bf Fig.~2.} Power deposition:~$t$-dependence \end{tabular}}}
\end{picture}}

\def\obriii{\unitlength=1bp\begin{picture}(220,190)(0,-15)
\put(10,0){\makebox(0,0)[bl]{\includegraphics[width=0.43
\textwidth]{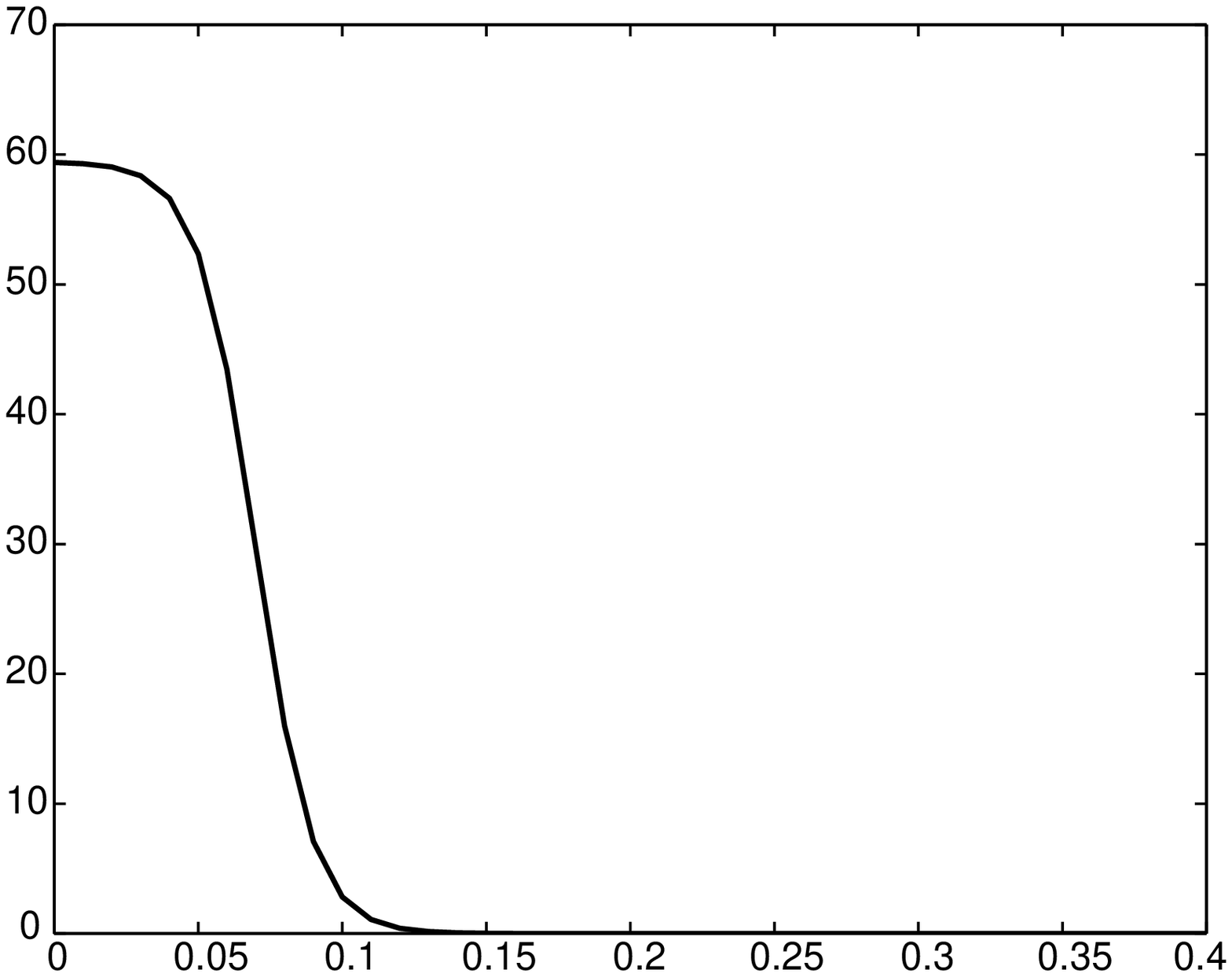}}}
\put(50,150){\makebox(0,0)[ct]{$q(x,t=0.5)$}}
\put(105,-5){\makebox(0,0)[ct]{$x$}}
\put(95,-20){\makebox(0,0)[ct]{
\begin{tabular}[t]{p{63mm}}
{\bf Fig.~3.} Power deposition:~$x$-dependence
\end{tabular}}}
\end{picture}}

\def\obrv{\unitlength=1bp\begin{picture}(220,190)(0,-15)
\put(10,0){\makebox(0,0)[bl]{\includegraphics[width=0.50
\textwidth]{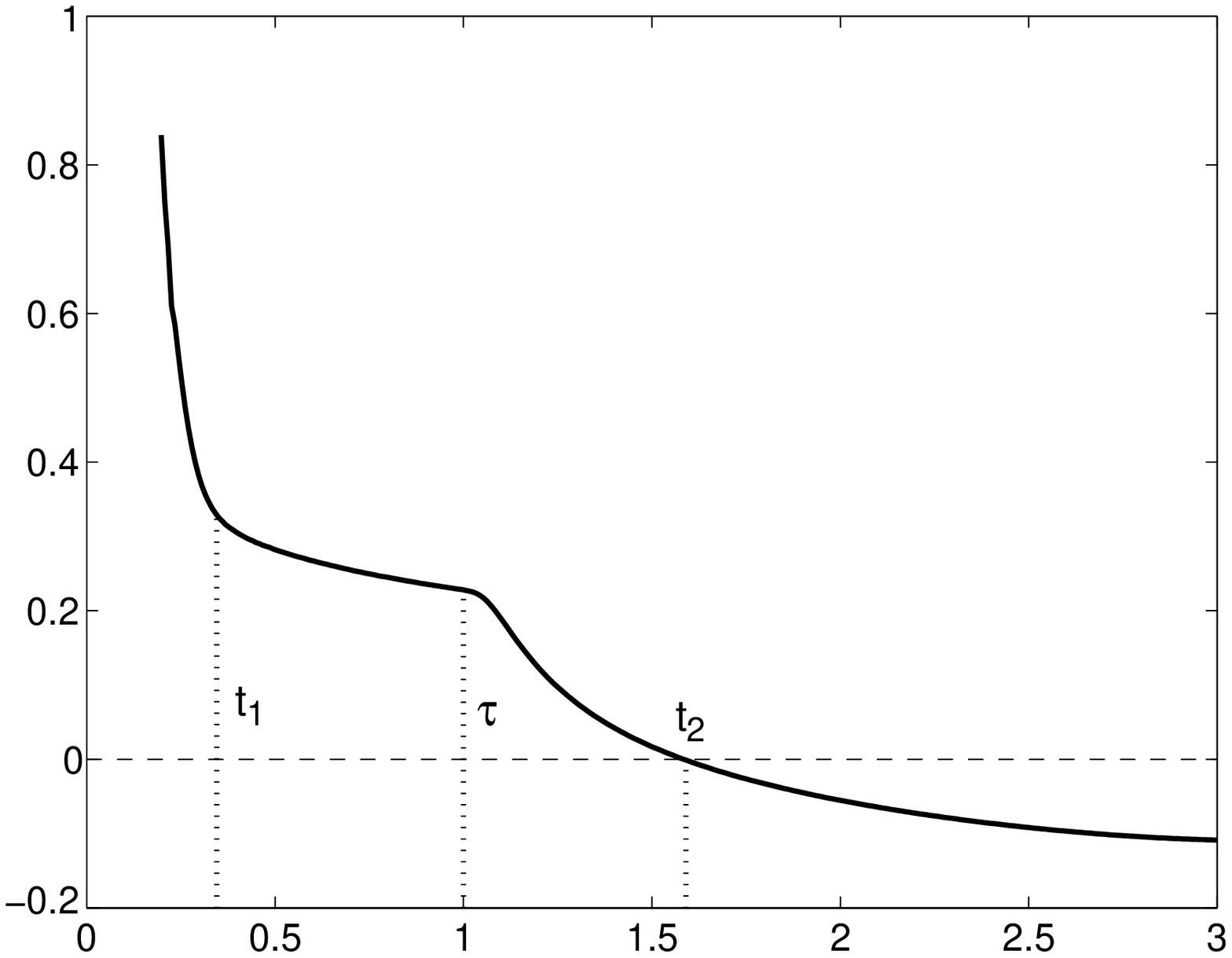}}}
\put(90,120){\makebox(0,0)[ct]{$\frac{d\xi}{dt}$}}
\put(105,-5){\makebox(0,0)[ct]{$t$}}
\put(130,-20){\makebox(0,0)[ct]{
\begin{tabular}[t]{p{80mm}}
{\bf Fig.~6.} Velocity of the boundary surface
\end{tabular}}}
\end{picture}}

\def\obrvii{\unitlength=1bp\begin{picture}(220,190)(0,-15)
\put(10,0){\makebox(0,0)[bl]{\includegraphics[width=0.50
\textwidth]{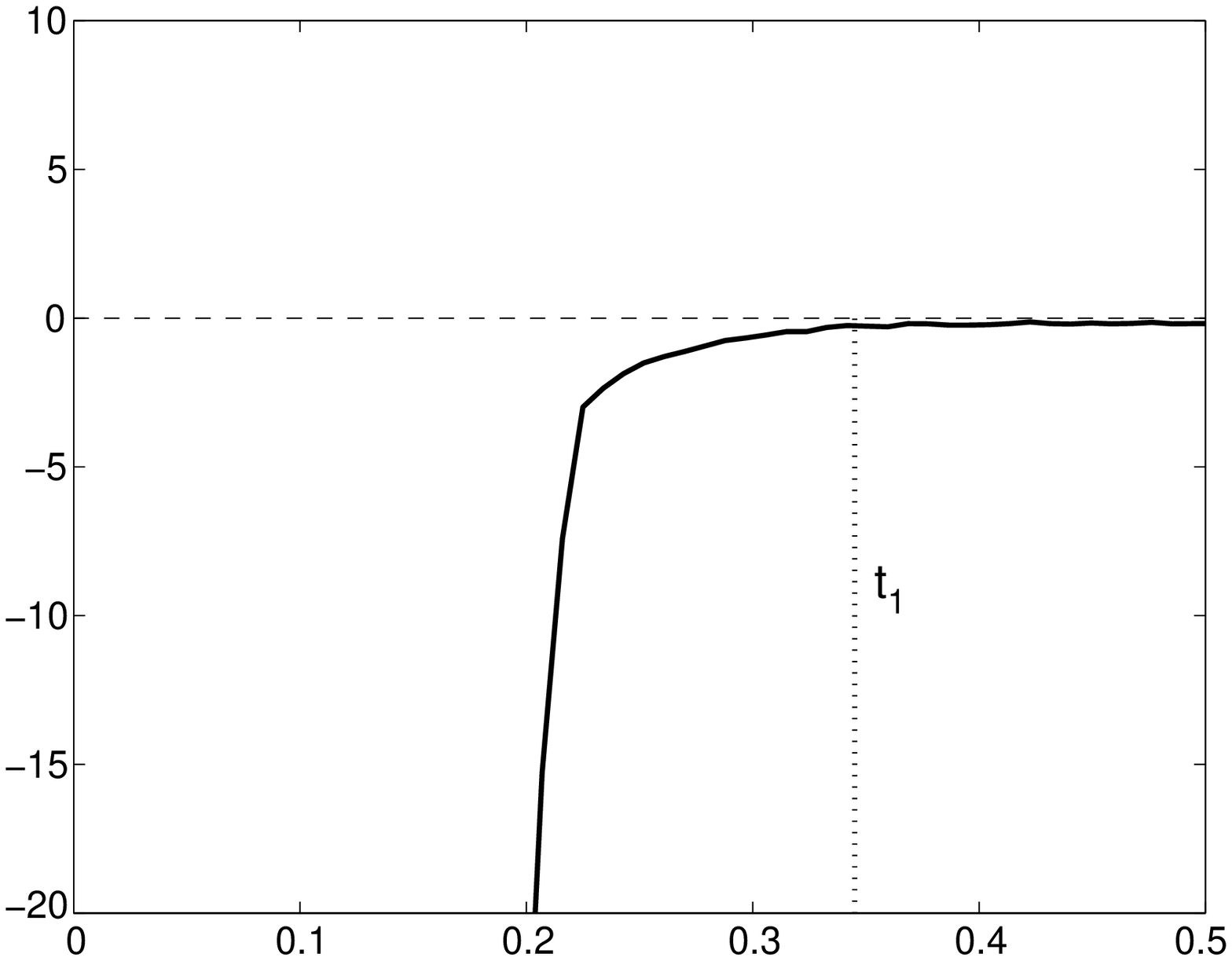}}} \put(60,130){\makebox(0,0)[ct]{$\phi(t)$}}
\put(105,-5){\makebox(0,0)[ct]{$t$}}
\put(168,-20){\makebox(0,0)[ct]{
\begin{tabular}[t]{p{60mm}}
{\bf Fig.~5. }
\end{tabular}}}
\end{picture}}

\def\obrx{\unitlength=1bp\begin{picture}(220,190)(0,-15)
\put(10,0){\makebox(0,0)[bl]{\includegraphics[width=0.43
\textwidth]{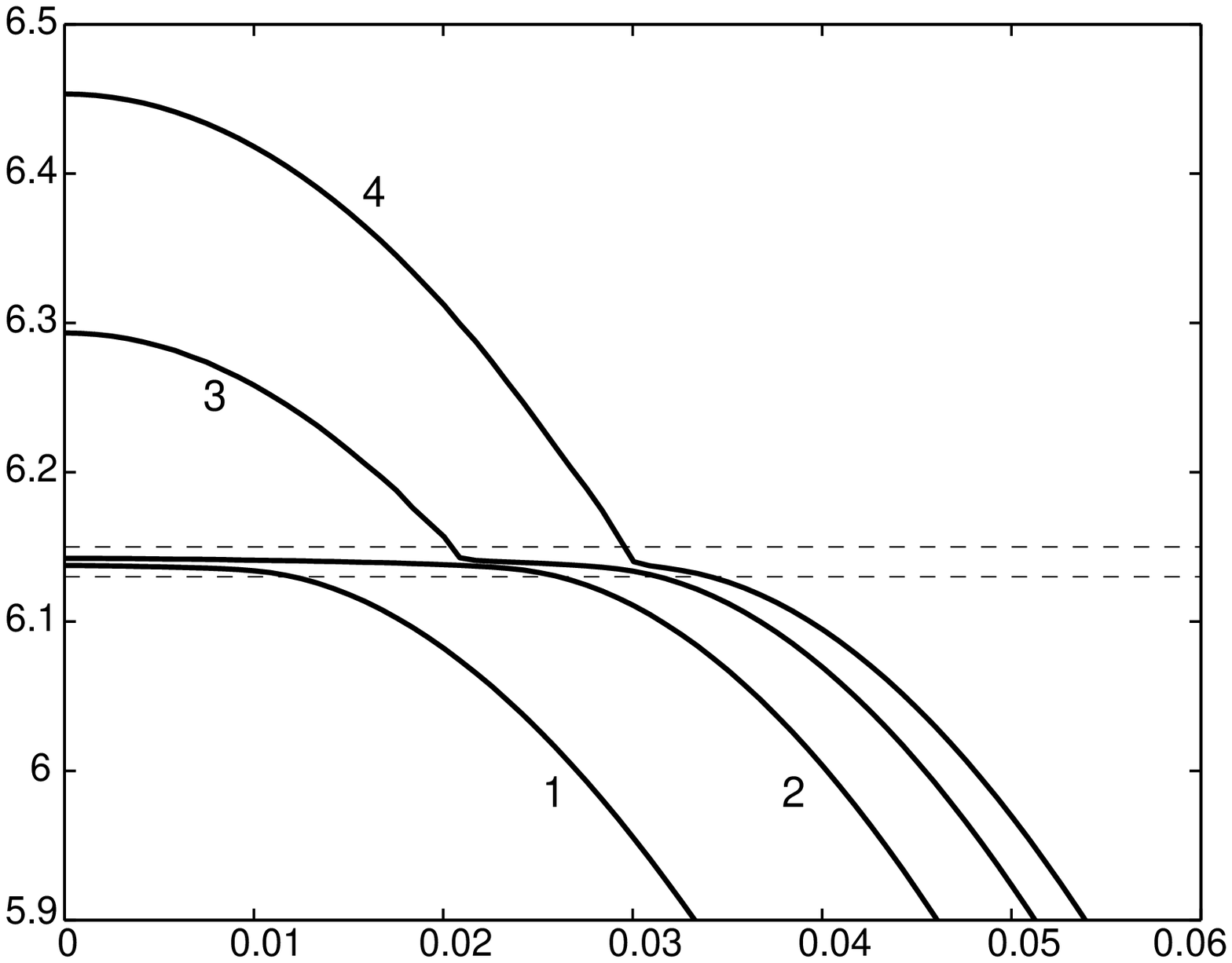}}} \put(100,125){\makebox(0,0)[ct]{$T(x)$}}
\put(105,-5){\makebox(0,0)[ct]{$x$}}
\put(95,-20){\makebox(0,0)[ct]{
\begin{tabular}[t]{p{63mm}}
{\bf Fig.~7.} Spatial temperature distribution for: $1)\;
t=0,162$; $2)\; t=0,186$; $3)\; t=0,204$; $4)\; t=0,216$
($\Delta=0,01$)
\end{tabular}}}
\end{picture}}

\def\obrxi{\unitlength=1bp\begin{picture}(220,190)(0,-15)
\put(10,0){\makebox(0,0)[bl]{\includegraphics[width=0.43
\textwidth]{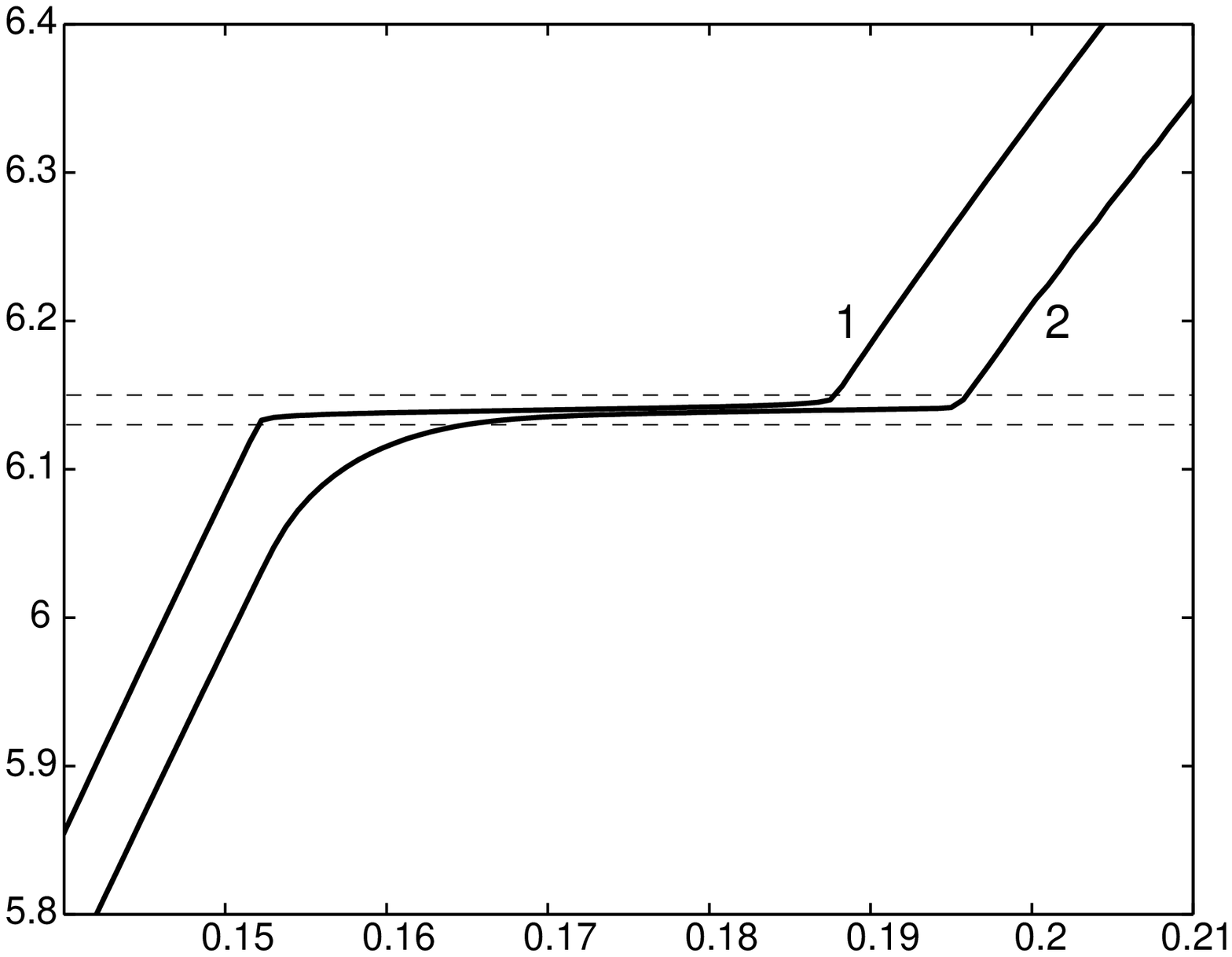}}} \put(100,125){\makebox(0,0)[ct]{$T(t)$}}
\put(105,-5){\makebox(0,0)[ct]{$t$}}
\put(94,-20){\makebox(0,0)[ct]{
\begin{tabular}[t]{p{63mm}}
{\bf Fig.~8.} Evolution of temperature for: $1)\; x=0$; $2)\; x
=0,04$ ($\Delta=0,01$)
\end{tabular}}}
\end{picture}}

\def\obrxiii{\unitlength=1bp\begin{picture}(330,280)(0,0)
\put(10,0){\makebox(0,0)[bl]{\includegraphics[width=0.85
\textwidth]{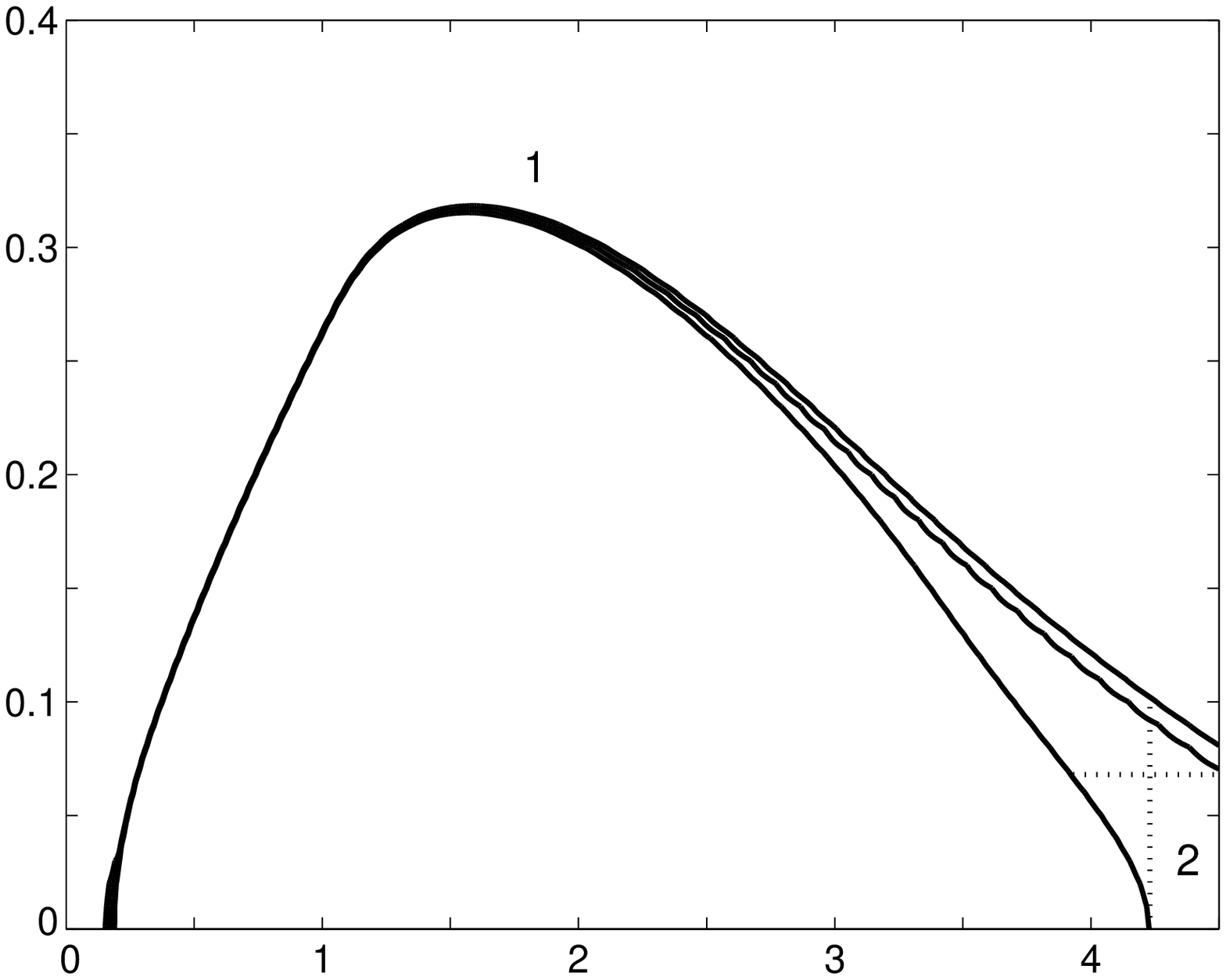}}} \put(60,175){\makebox(0,0)[ct]{$\xi(t)$}}
\put(140,-5){\makebox(0,0)[ct]{$t$}}
\put(160,-20){\makebox(0,0)[ct]{
\begin{tabular}[t]{p{100mm}}
{\bf Fig.~9.} Time dependence of the interphase coordinate. Here a
curve in the middle corresponds to the fusion  temperature equals
$T^*$. For the upper and lower curves, the fusion temperature is
chosen to be at $T^* - \Delta/2$ and $T^* + \Delta/2$, accordingly
($\Delta=0,01$)

\end{tabular}}}
\end{picture}}

\def\obrxiv{\unitlength=1bp\begin{picture}(0,0)(0,0)
\put(-107,166){\makebox(0,0)[bl]{\includegraphics[width=0.32
\textwidth]{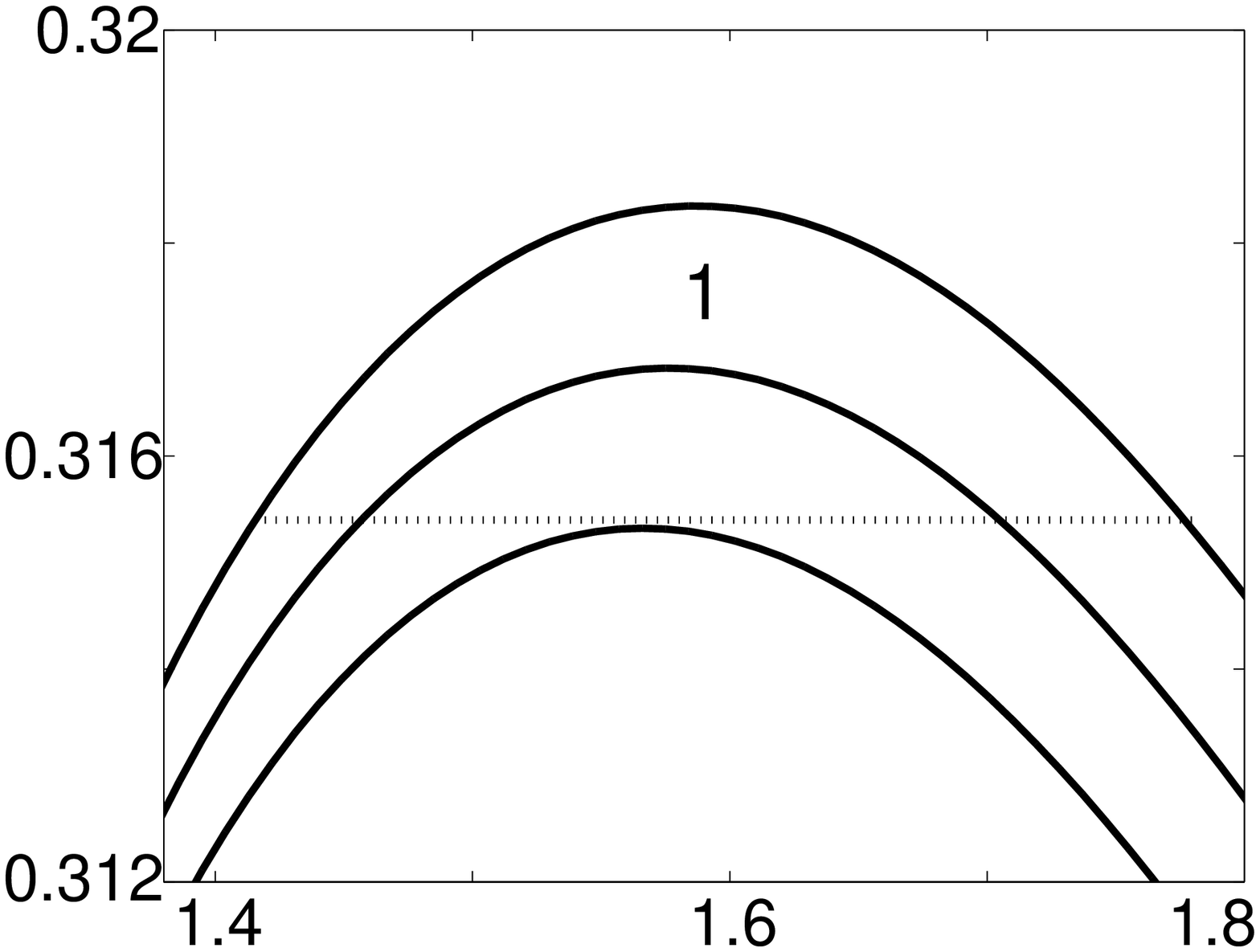}}}
\end{picture}}

\def\obrxv{\unitlength=1bp\begin{picture}(280,200)(0,-15)
\put(10,0){\makebox(0,0)[bl]{\includegraphics[width=0.65
\textwidth]{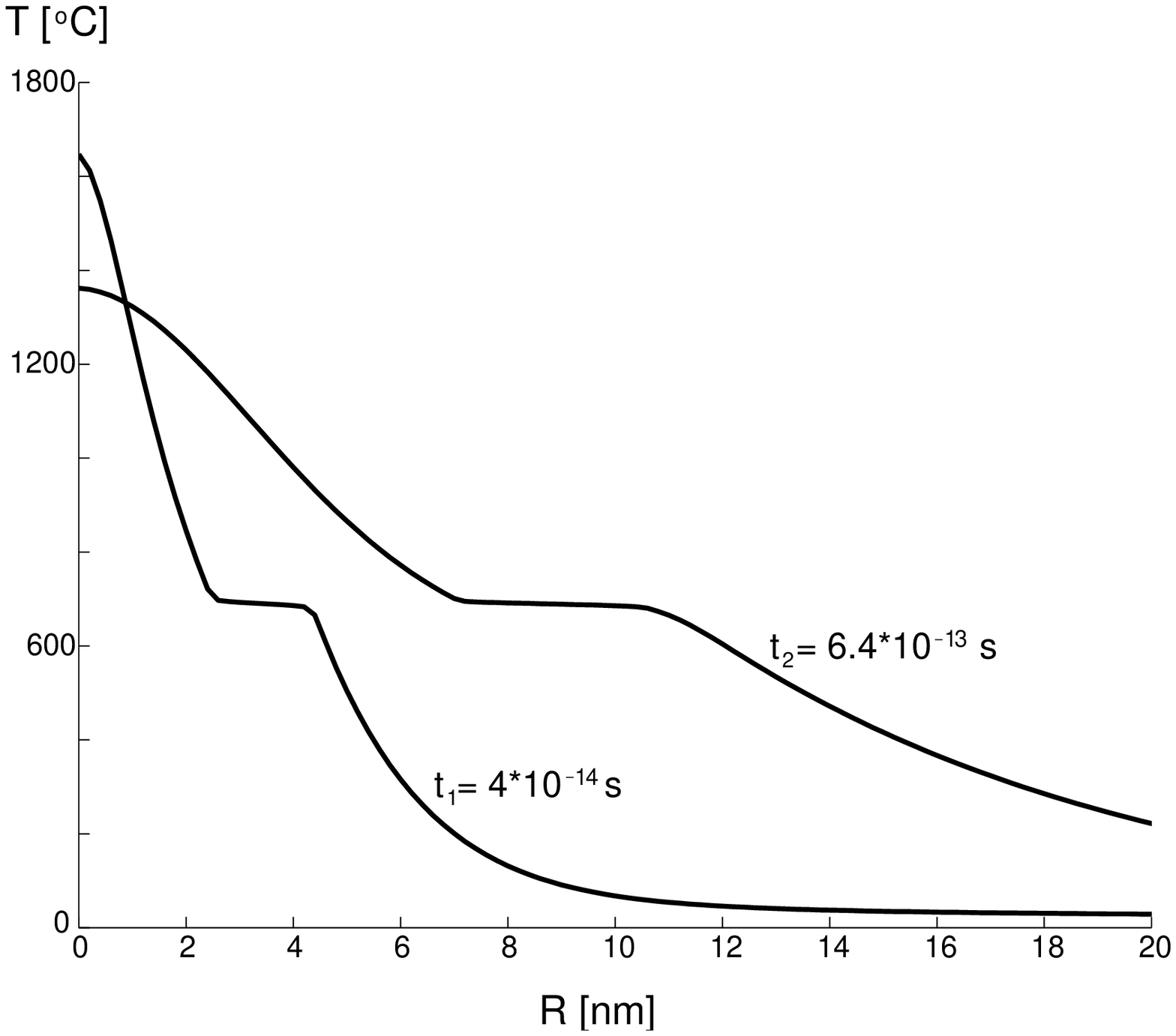}}} \put(145,-20){\makebox(0,0)[ct]{
\begin{tabular}[t]{p{95mm}}
{\bf Fig.~10.} The radial distribution of the lattice temperature
$T_i$ for two different moments of time
\end{tabular}}}
\end{picture}}

\begin{document}

\begin{center}
{\Large \bf Stefan problem and beyond}\\
\vspace{5mm}

{\bf  B.F.~Kostenko, J.~Pribi\v{s}, I.V.~Puzynin}\\
 {\it Joint
Institute for Nuclear Research, Dubna \\
 141980 Moscow region, Russia}
\end{center}

\vspace*{10mm}


\vspace*{0.5cm} \centerline{\large \bf Abstract} We claim that the
celebrated Stefan condition on the moving interphase,  accepted in
mathematical physics,  can not be imposed if energy sources are
spatially distributed in the volume. A method  based on Tikhonov
and Samarskii ideas  for numerical solution of the problem is
developed.
Mathematical modelling of energy relaxation of some processes
useful in modern ion beam  technologies is fulfilled. Necessity of
taking into account effects completely outside the Stefan
formulation is  demonstrated.

\vspace{15mm}

\vspace{15mm}

The address for  correspondence: \vspace{5mm}\\
 {\bf
Dr. B.F. Kostenko\\
Laboratory of Information Technologies\\
Joint Institute for Nuclear Research\\
141980, Dubna, Moscow region\\
Russia\\
\vspace{1cm}

E-mail: kostenko@jinr.ru\\
Tel.: +007-096-21-64-069; fax: +007-096-21-65-145 \\
}


\newpage
\section{Introduction}

The  Stefan problem concerns  solid-liquid or liquid-vapor phase
transitions when moving {\bf unknown beforehand} surface $S$ of
phase transition is formed (see, e.g., \cite{Samarskii1}). In
fact, a formulation of the Stefan problem was given for the first
time by G.~Lame and B.P.~Clapeiron in 1831 for a particular case
of equal temperature of liquid and crystalline phases \cite{Lame}.
In 1889 J.~Stefan published four papers devoted to the subject (in
particular, to the description of soil freezing) in which the
problem was formulated in a general form \cite{Stefan}. According
to it, for the interphase the following condition
\begin{equation}
K_{sol} \frac{\partial T(x_S + 0, t)}{\partial x} - K_{liq}
\frac{\partial T(x_S - 0, t)}{\partial x}  = L \rho_{sol} V_S,
\end{equation}
defining   Stefan's problem, has been  suggested. Here $V_S = d
\xi_S / d t$ is the velocity of the boundary surface $S$,
$K_{sol}$ and $K_{liq}$  are thermal   conductivities  of material
for solid and liquid phases,  $L$ and  $\rho_{sol}$ are the
melting heat and density, correspondingly. Condition (1) has a
clear physical meaning. Indeed, according to the Fourier law, heat
flow $j$ is proportional to the temperature gradient,
$$
{\bf j}\;=\;-K\; grad \; T.
$$
Therefore, the left-hand side of (1) is the heat absorbed in the
unit of area  per the unit of time. The expression in the
right-hand side is the heat connected with  freezing or melting of
material crossed per the unit of time by the  the unit of area.

A complete mathematical formulation of the  Stefan problem
includes, besides (1), a condition of  continuity on the surface
$S$ separating solid and liquid phases
\begin{equation}
T|_S = T^* ,
\end{equation}
where $T^*$ denotes  temperature of the phase  transition, and the
energy conservation law
$$
\rho C \;  \frac{\partial T }{\partial t} = - div \;{\bf j} +
q({\bf x}, t).
$$
Here $q({\bf x}, t)$ represents the power of external heat
sources, $C$ is  the specific heat. In the original Stefan papers
$q({\bf x}, t) \equiv 0,$ so that the whole heat transfer has been
considered  to be a consequence of the temperature gradient inside
the medium.

If one also specifies initial and boundary conditions, the Stefan
problem can be solved more often approximately, but sometimes
exactly. Particular examples of suitable boundary conditions are
considered below.

Relations (1) and (2) are usually used   in numerical algorithms
explicitly. Another  approach was suggested by A.N.~Tikhonov and
A.A.~Samarskii in 1953 \cite{Samarskii2}. According to it,
conditions (1) and (2) are included themselves  into the energy
conservation equation to obtain generalized formulation of the
Stefan problem in the form
\begin{equation}
(\rho C + L \; \delta (T-T^*)) \left( \frac{\partial T }{\partial
t} + {\bf v} \; grad \; T\right)= div(K \; grad \; T) + q({\bf x},
t),
\end{equation}
where the term $L \; \delta (T-T^*)\;
\partial T / \partial t$ describes the additional heat input
expended on the phase transformation, ${\bf v} \; grad \; T$ takes
into account possible temperature change due to convection
(hereafter  we ignore it for simplicity).  The main idea of this
approach is  quite clear, too. Namely, it is suggested to treat
the heat of fusion $L $  as an additional component of the thermal
capacity $\rho C$ which gives  contribution only at the point of
phase transition.

Lately Samarskii  and  his followers  have turned this  idea into
effective numerical algorithms (see, e.g. \cite{Samarskii3}). But
even in those papers equation (3) is considered only  as a
corollary of the condition (1). For example, it was derived in
\cite{Samarskii1} by substituting expression $L \; \delta
(T-T^*)\;
\partial T /
\partial t$ instead of the term $L \; \delta ( x -  \xi_S (t))\; V_S $,
which is assumed to be included in the heat equation to
account for the heat absorption on   2-dimensional interface $S$.

The purpose of this paper is to show that the condition (3)
supplies us with a more powerful description of phase transitions,
that may be used even in the case when (1) and (2) are not
applicable.

\section{Heuristic arguments}

As it was mention above, the possibility of  solving the classical
Stefan problem by making use of condition (3) has been
demonstrated by Samarskii  and his co-authors. Therefore, we only
consider   an example when (3) is applicable  and (1), (2) are
not. To this end let us study  the following  problem:
\begin{equation}
(\rho C + L \; \delta (T-T^*)) \; \frac{\partial T }{\partial t}
div(k \; grad \; T) + q(t),
\end{equation}
$$
T({\bf x}, 0) = T_0 < T^* ,
$$
where all parameters of (4) are suggested to be independent of
$\bf x$. Due to the spatial uniformity, it is evident that  the
condition
$$
grad \; T = 0
$$
holds on the solutions of (4).  In this case Eq. (4) is reduced to
an ordinary differential one
\begin{equation}
(\rho C + L \; \delta (T-T^*)) \; \frac{d T }{d t} = q(t)
\end{equation}
with the initial condition
$$
T(0) = T_0 .
$$
Integrating both sides of (5) over $t$  just near the phase
transition temperature $T^*$, one obtains
\begin{equation}
\int_{T^* -0}^{T^* +0} (\rho C + L \; \delta (T-T^*)) \; d T  =
\int_{t}^{t+\delta t} q(t) \; dt,
\end{equation}
where $\delta t$ is a time necessary for the phase transition. It
is evident from (6) that
\begin{equation}
\delta t \geq \frac{L}{Q},
\end{equation}
where $Q$ is the maximum value of $q(t)$ in the interval $(t, t+
\delta t)$. The  inequality (7) means that the phase transition at
a fixed spatial point lasts  {\bf a finite,  distinct from zero,
time.}

This simple example  shows  something completely different from
the Stefan description  of the phase transition. Let us examine it
carefully.
\begin{enumerate}
\item
First of all, instead of gradual warming (or cooling)  up  the
pattern due to the influence of one of its boundaries, here we
have an uniformly heated layer. Therefore, creation of 2-D surface
$S (y,z) $ separating the solid and liquid phases in $x$  is
evidently impossible due to a  total equivalence of all spatial
points $x$.
\item
One can also expect that finiteness of the phase transition time,
$\delta t$, forces all points within {\bf a spatial   layer of
nonzero thickness}  to be at the same temperature $T^*$. This is
expected even in the case when the power deposition $q(x,t$),
unlike in the example considered, is spatially irregular\footnote{
Indeed, let material  has just reached the temperature $T^*$ at
some point $\bf x$ and now begins  receiving  its portion of the
heat necessary for melting. Then  another adjacent point $\bf x$
$+$ $\Delta \bf x$, which attained the melting temperature merely
a little earlier, can be still in the state of heat receiving and,
therefore, must have {\bf the same temperature $\bf T^*$} (see
Fig.1).}.
\end{enumerate}

\centerline{\quad\qquad\obrxxi\obrxxii}\vspace{25mm}

If we consider $\delta$-function in (3) as a limit of a bounded
function $D(T-T^*)$ localized in the vicinity of $T=T^* $, then
the possibility to obtain the  solution shown in Fig. 1(a), or
analytically,
$$
  \frac{\partial T}{\partial t} \longrightarrow 0,  \qquad
grad \; T \longrightarrow 0 ,
$$
follows from the indefiniteness
$$
D(T-T^*)\; \frac{\partial T}{\partial t} \; \longrightarrow \;
\infty \; \cdot \;   0,
$$
springing up in the left-hand side of (3). Clearly, that this
indefiniteness can take a finite value and compensate in a space
region with nonzero thickness  the spatially distributed source
$q({\bf x}, t)$ which contributes to the right-hand side of (3).
This, of course, is no more true if the external sources are
absent and   heat enters the pattern only through its boundary.

In a general case, one can expect existence  of {\bf two jumps}
for spatial derivatives   of  temperature on the boundaries $S$ of
the {\bf volume} $V_{\;T^*}$ with $T=T^*$, instead of one for the
classical Stefan problem,  but the condition (1) is hardly met for
any of them (see Fig.~1(a), where intersection of the boundary $S$
by (x,y)-plane in points 1, 2, 3 and 4 is seen). Indeed, to prove
the existence of two jumps --- one from the side of the solid  and
another from the side of the  melted phase --- it is sufficient
only to show that the spatial derivative on the surface $S$,
{\bf taken externally}, is not equal to zero. The co-ordinates of
the boundary $ \vec \xi (t)$ can be found as a solution of an
equation
$$
T({\bf x},t) - T^* = 0 ,
$$
where $T({\bf x},t)$ is the solution of the heat equation (3)
outside the volume $V_{\;T^*}$. Taking  the total temporal
derivative, one obtains
$$
\frac{\partial T}{\partial t} + grad \; T \;\cdot \; \frac{d \vec
\xi}{dt} = 0.
$$
Thus $grad \; T = 0$ automatically implies $ \partial T / \partial
t = 0.$ It is evident  that such conditions are impossible if the
external sources are not adjusted specially  to stabilize  the
temperature in the infinitesimal layers adjacent to the volume
$V_{\;T^*}$ just before  and just after the phase transition.

\section{Beam induced phase transitions}

To verify the conclusions which  we have just come to, let us
study {\bf numerically} the dynamics of phase transition induced
by a short powerful ion beam in solids. At present this technology
is really used for modification of surface layers to create new
materials with unique physical and chemical properties (see, e.g.
\cite{Bleikher}). The process is underlain by the  equation for
heat transfer which we  discussed in the previous sections:
\begin{equation}
\rho (T) c(T)  \frac{\partial T}{\partial t} = \frac{\partial
}{\partial x} \left( k(T) \frac{\partial T}{\partial x} \right) +
q.
\end{equation}
The initial and boundary conditions could be taken in the form:
$$
T(x,0) = T_0, \qquad  \frac{\partial T(0, t)}{\partial x}=
\frac{\partial T(l_0, t)}{\partial x} = 0.
$$

Let us consider, for definiteness, an iron pattern, which thermal
properties are described in popular reference books, and choose
dimensionless ({\bf DL} for brevity) variables  $$T:=T/T_0 , \;
x:=x/l_0 , \; t:= t/\tau
$$ as follows:

$ T_0 = 293 \; K, \hspace{0.5cm} $ $l_0 = 10^{-5}\; m \;({\rm the
\; pattern \; thickness}),$

$\tau = 3\; 10^{-7}\; s \; ({\rm duration \; of \; ion \; beam \;
pulse\; from\; an\; accelerator}). $

For DL power deposition $q$ we  take a simple model, shown in Fig.
2 and 3,

\centerline{\quad\qquad\obrii\obriii} \vspace{15mm}

\noindent with analytical representation
$$
q(x,t) = Q \; q_1 (x) q_2 (t) ,
$$
where
$$
q_i (z) = \frac{1}{1+ \exp \mu_i (z-z_i )}
$$
and $Q$ describe the total DL energy brought into the pattern
(here $Q= 59.44, \; x_1 =0.07, \; t_1 =1,\; \mu_i =100 $). For
simplicity, we neglect in (8) a small difference between physical
parameters for the solid and liquid phases.

Now, using of  the general idea due to Tikhonov and Samarskii
\cite{Samarskii2}, we assume  an expression:
$$
\rho (T) c(T) = 1 +  \lambda \delta (T- T^* , \Delta )
$$
for DL specific heat, where $\lambda$ denotes the DL heat of
fusion and $\delta (T- T^* , \Delta )$ is an approximate
$\delta$-function, smoothed with the help of the Gaussian
distribution of width $\Delta$ (see Fig. 4)\footnote{There were
other methods of smoothing in original papers  by Samarskii et al.
They used regularization on the space grid.}.

Now Eq. (8) can be solved numerically on the space-time grid $x$
and $t$ with steps $h_x$ and $h_t$, within intervals $x \in (0,1),
\; t \in (0, t_{max} )$:

$$
x_j = h_x \cdot j, \; j=0,\dots,n_x, \; h_x=1/n_x,
$$
$$
t_k = h_t \cdot k ,\; k=0,\dots,n_t, \; h_t=t_{max}/n_t,
$$
where $n_x$ and $n_t$ are numbers of partitions.

\centerline{\obri} \vspace{15mm}

The following difference scheme with weights $\gamma$ was
implemented (see \cite{Kalitkin} for details):

\begin{equation}
e_j^k \frac{T_j^{k+1}-T_j^k} {h_t}=k_0 \left[ \gamma
\frac{T_{j+1}^{k+1}-2T_{j}^{k+1}+T_{j-1}^{k+1}}{h_x^2}+
(1-\gamma)\frac{T_{j+1}^{k}-2T_j^{k}+T_{j-1}^{k}}{h_x^2}
\right]+q_j^{k+\frac{1}{2}},
\end{equation}

\noindent where
$$
T_j^k=T(x_j,t_k), \quad e_j^k=\rho(T_j^k)c(T_j^k), \quad
q_j^{k+\frac{1}{2}}=q(x_j,t_k+\frac{h_t}{2}),
$$
and the upper index numerates different moments of time (time
``levels''), the lower one specifies a set of spatial
co-ordinates. The scheme is absolutely convergent at $\gamma =
0.5$ and possesses  the second-order accuracy for both variables.

From initial condition $T(x,0)=T_0$, values $T_j^0$
$(j=0,\dots,n_x)$ on a zero time level are known. The boundary
conditions
$$
\frac{T_{1}^k-T_{-1}^k}{2h_x}=\frac{T_{n_x+1}^k-T_{n_x-1}^k}{2h_x}=
0, \quad k=1,\dots,n_t
$$
allow one to introduce  symmetric points $x_{-1}=-h_x$ and
$x_{n_x+1}=1+h_x$ with  appropriate values $T_{-1}^k=T_1^k$ and
$T_{n_x+1}^k = T_{n_x-1}^k$ respectively. So, we can use the Eq.
(9) in  points $x_0$ and $x_{n_x}$. Using initial and boundary
conditions, we obtain a system of $n_x$ linear algebraical
equations with the same number of variables. Thus, under the
accepted approximation, we reduced the partial differential
equation (8) to  system (9) of linear algebraic equations. The
matrix of this system is tridiagonal and after its
solution\footnote{Recursive relations for determining the solution
of algebraic problem (9) comprise the well-known sweep method,
called  also forward-backward or Thomas algorithm
\cite{Samarskii1}. } we obtain value $T_j^1$ $(j=0,\dots,n_x)$ at
the first time level. Repeating this process, values $T_j^k$ on
every time level $k$ are computed.

The result of straightforward verification of the Stefan condition
(1)  is shown in Fig. 5, where the function
\begin{equation}
\phi(t)=k\left.\left(\frac{\partial T}{\partial x} \right|_{T_A}-
\left.\frac{\partial T}{\partial x}\right|_{T_B}\right)-\lambda
\frac{d \xi}{dt}
\end{equation}
is depicted.

\centerline{\obrvii}\vspace{15mm}

The left and right points, in which the spatial derivatives of
temperature were taken in (10), are shown in Fig. 4. They define a
spatial layer which nearly the whole fusion energy  is absorbed
within. From Fig. 5 one can see that condition (1) is satisfied
indeed, but only {\bf after a characteristic relaxation time $\bf
t_1$ has elapsed.} The physical meaning of $t_1$ is clear from
Fig. 6. Namely,  it corresponds to the transition from a rapid to
slow motion of the exterior interphase surface. In the  case when
boundary motion is rapid,  the heat necessary for fusion is
brought into the melting layer {\bf directly from the external
source} $q(x,t).$ The slow motion corresponds to the ordinary
Stefan mode when the process is controlled mainly by the heat
entered into the layer through its boundary.

\centerline{\obrv}\vspace{15mm}

It is also seen from Fig. 6 that transition to the Stefan mode
takes place earlier than the external source to be totally turned
off:
$$
t_1 < \tau .
$$
Time $t_2 $ shown in Fig. 6  denotes a moment when the thickness
of the melted material begins to diminish due to heat  escape into
a more cooler solid phase.

Fig. 7 and 8 also confirm the conclusions which we have come to in
the previous section. Formation of the ``tableland'' (whose height
corresponds to the  fusion temperature)  for spatial temperature
distribution is distinctly seen in Fig. 7. The narrow strip
restricted  by two dashed lines in Figs. 7 and 8 exhibits the
width of the smoothed $\delta$-function. We believe that existence
of {\bf two} breaks for the spatial derivative is masked in Fig. 7
with this $\delta$-function smearing.
Fig. 8 demonstrates a temperature evolution   for two divorced
spatial points. One can make sure that the above mentioned time
interval corresponding to the {\bf same temperature} at the {\bf
different spatial  points} really exists. It is evident that such
a behavior of temperature  has nothing to do with the traditional
description in the framework of (1) and (2).

\centerline{\quad\qquad\obrx\obrxi}\vspace{15mm}

\vspace{1cm}

Fig. 9 shows a time-dependence of the interphase coordinate.
Numbers 1 and 2  denote the regions where verification of the
Stefan condition (1) is impossible due to $\bf \Delta$-{\bf
instability}. It means that small variations of fusion temperature
value, $T^*$, lead to a drastic change of interphase position (see
dotted lines in Fig.~9).

\centerline{\obrxiii \obrxiv}

\vspace{30mm}

\section{Track formation in solids}

The next example  demonstrating the preference for the
$\delta$-function approach is connected with the problem of track
formation in solids. In fact, at present nobody knows with
certainty the main mechanism responsible for these processes.
Furthermore, it seems like  the universal model explaining all of
them does not exist and different materials behave differently
under  heavy ion attack. Here we assume the so-called thermal
spike model based on the following system of two coupled nonlinear
differential equations (see, e.g. \cite{Toulemonde} and references
therein):

\begin{equation}
\rho C_e (T_e  )\frac{\partial T_e }{\partial t}=\frac{1}{r}
\frac{\partial}{\partial r} \left[r K_e (T_e ) \frac{\partial
T_e}{\partial r}\right] - g\cdot (T_e -T_i )+q(r,t),
\end{equation}

\begin{equation}
\rho C_i(T_i )\frac{\partial T_i }{\partial t}=\frac{1}{r}
\frac{\partial}{\partial r} \left[r K_i (T_i )\frac{\partial T_i
}{\partial r}\right] + g\cdot (T_e-T_i ),
\end{equation}
where $T_e$ and $T_i$ are electrons and lattice temperatures,
respectively, ~$C_e$,~$C_i$ and ~$K_e,K_i$    specific heat and
thermal conductivity for the electronic system and lattice,
~$\rho$ is the material density, ~$g$  the electron-atom coupling,
~$q(r,t)$  the power brought on the electronic system,  $r$ the
radius in cylindrical geometry with the ion path as the axis. One
can see   that electrons receive  their energy directly from the
external source $q(r,t)$ which takes into account  ion energy loss
in electron gas. The characteristic duration of source activity is
usually  in the range $10^{-15} - 5\times 10^{-15}$ s. According
to (12), atoms are heated due to electron-atom coupling
represented by the term $g\cdot(T_e-T_i )$. Nuclear interaction of
atoms  with the projectile ion is relatively small and, therefore,
can be neglected. It is clear that coupling is the most effective
at the beginning of the relaxation process when $T_e \gg T_i$ and
$g\cdot(T_e-T_i ) \simeq g T_e$.

The initial  conditions can be chosen in a form
$$
T_e (r,0) = T_i (r,0) = T_0 ,
$$
and the boundary ones\footnote{ One should take into account that
there is no heat transfer at the center of track.} can be taken as
$$
\left( \frac{\partial T_e}{\partial r} \right) _{r =0} =
\left(\frac{\partial T_i}{\partial r} \right) _{r =0} =0, \qquad
T_e (r_{max}, t) = T_i (r_{max}, t) = T_0 ,
$$
where  $ r_{max}$ was taken  of order $10^{-5}\; cm$.

The thermal spike model explains track formation  as a structural
transition of lattice due to its warming-up  and melting with
subsequent   quenching. These processes are usually accompanied
with disorder creation   in the lattice. Indeed, rapid quenching
leads  to a ``conservation'' of atoms' random places that were in
the melted material just before cooling. For amorphous materials,
which are characterized by high disorder of atoms' positions  and
small values of thermal conductivity, quenching, quite the
contrary, leads to putting atoms' places in order. But in either
case, structural modifications are observed in the microscope as
an ion trace in solid.

Besides thermal spike, one may assume the ion spike as well, when
the track is formed due to the electrostatic repulsion of ionized
atoms.
The main  reason justifying our utilization of system (11), (12)
is an agreement of nuclear track radii, calculated in this
framework, with the experimental data \cite{Toulemonde}.
The total formulation  of the model includes many physical
details, such as a description of the source  $q(x,t)$,  and is
outside the scope of this publication. Here we only touch  some
problems concerning the main topic of the paper.

A numerical algorithm similar to that described above has been
elaborated for numerical solving system (11), (12). The radial
distribution of the lattice temperature $T_i $ around the path of
$\bf Pb$ in amorphous $\bf Ge$ at kinetic energy  of impinging
ions  of about $110 $ MeV is shown in Fig. 10 (for two different
moments of time). One can see the typical ``tablelands'' similar
to those discussed in the previous chapter and which could not be
obtained in  frames of the classical Stefan approach. Our
calculations show that the ``tableland'' exists here only during a
transitory time $t_1$, when the material is under a strong
exposure of the source $g\cdot(T_e-T_i ) \simeq g\; T_e$. It is
shorter than $\tau= \varrho C_e /g \approx 10^{-12}$ s (see Fig.
10), where $\tau$ is electron-atom relaxation time\footnote{The
formula for $\tau$ estimation follows from Eq.~(11).},
characterizing duration of source activity in (12) (compare this
conclusion with data presented in Fig. 6).

\vspace{10cm}
\centerline{\obrxv}\vspace{15mm}


It is interesting to note that there is a real, met in the nature,
``regularization'' of $\delta$-function  analogous to that
implemented  in this paper. For materials with a  complex
molecular structure (high temperature superconductors, biological
molecules, alloys etc.), the melting temperature is not fixed but,
instead, smeared within a characteristic interval where atom bonds
of different type are gradually destroyed with temperature
increase. In this case the only possible approach to the problem
should be based on the condition (3). An approximate
$\delta$-function, analogous to that  shown in Fig.~4, can be
extracted here  directly from the experiment. In \cite{Goncharov},
a model based on the smeared $\delta$-function approach and Eqs.
(11), (12) was used for computation of effective electron-atom
relaxation time $\tau$ in a high temperature superconductor. The
established $\tau$ turned out to be in a good agrement with
experimentally observed values.

\section{Conclusion}

To the best of our knowledge, the peculiarities  of phase
transition dynamics, we discussed in this paper, have never been
considered explicitly in mathematical physics.  This fact may be
explained partially by the necessity to use {\bf very powerful}
spatially distributed external sources of heat, in order the above
mentioned effects to be urgent. Such sources were hardly available
for industrial applications  not long ago. However, the examples
which  given above are likely evidences of the fact that such
sources,  ``interfering'' in the thermal conductive processes, are
integral parts of {\bf all} most recent ion beam technologies. The
numerical investigations, which have been undertaken,  show that
the $\delta$-function approach to phase transitions is a suitable
instrument to tackle these problems, though the authors of this
idea have never used it in such a context.

Within networks of scientific papers devoted to the interphase
motion problem, one can distinguish in retrospect  the following
logical order: formulation of the Stefan problem (Lame and
Clapeiron; Stefan) $\longrightarrow$  application of the
$\delta$-function approach for equivalent representation of it
(Tikhonov and Samarskii) $\longrightarrow$ numerical
implementations of the idea (Samarskii with co-authors; this
review) $\longrightarrow$ description of materials with  interval
distributed  fusion temperature  (a natural physical
interpretation of the previous step). Here we have shown that it
is more expedient to {\bf turn over} this order and  take its last
element as the basis for solving  both the classical Stefan
problem and a more general one including spatially distributed
sources. In other words, both of these solutions could be
considered as an idealized limiting case $\Delta \rightarrow 0$ of
a natural physical point of view that none of phase transitions
take place at the {\bf exactly} defined value of fusion
temperature. A peculiarity of  a new, found in this paper,
solution to Eq. (3) is its ``tableland'' behavior seen in Figs.
1(a), 7, 8 and 10. Such  solutions could never be obtained in the
framework of the classical Stefan formulation (compare with Fig.
1(b)).

We would like to express our gratitude to professor
M.~Toulemonde from CIRIL (Caen) for interest and  useful
discussion. The authors are grateful to doctors E.~Airjan,
I.~Amirkhanov, I.N.~Goncharov and T.P.~Puzynina from JINR (Dubna)
for stimulating discussions.

This investigation has been supported in part by the Russian
Foundation  for Basic  Research, project No. 02-01-00606.

\end{document}